\pdfoutput=1
\documentclass[11pt]{article}

\usepackage[final]{acl}
\usepackage{tcolorbox}
\usepackage{times}
\usepackage{latexsym}
\usepackage{colortbl}      % 允许单元格着色 (\rowcolor, \cellcolor)
\usepackage{xcolor}        % 允许自定义颜色
\usepackage{amsmath}
\usepackage{amsmath, amssymb}
\usepackage{makecell}
\usepackage{enumitem}
\usepackage[T1]{fontenc}
\usepackage{algorithm}
\usepackage{algpseudocode}
\usepackage[utf8]{inputenc}

\usepackage{microtype}
\usepackage{multirow}
\usepackage{booktabs}
\usepackage{inconsolata}

\usepackage{graphicx}

\usepackage{fancyhdr}
\title{Secure Tug-of-War (SecTOW): Iterative Defense-Attack Training with Reinforcement Learning for Multimodal Model Security}

\author{
 \textbf{Muzhi Dai\textsuperscript{1}\thanks{Work done during an internship at TeleAI}
\thanks{Equal Contribution}},
 \textbf{Shixuan Liu\textsuperscript{1}\footnotemark[2]},
 \textbf{Zhiyuan Zhao\textsuperscript{1}\thanks{Corresponding Author}},
 \textbf{Junyu Gao\textsuperscript{1, 2}},
 \\
 \textbf{Hao Sun\textsuperscript{1}},
 \textbf{Xuelong Li\textsuperscript{1}\footnotemark[3]}
\\
 \textsuperscript{1}Institute of Artificial Intelligence (TeleAI), China Telecom, China
 \\
 \textsuperscript{2} Northwestern Polytechnical University, China
}

\begin{document}
\maketitle
\begin{abstract}
The rapid advancement of multimodal large language models (MLLMs) has led to breakthroughs in various applications, yet their security remains a critical challenge. One pressing issue involves unsafe image-query pairs-jailbreak inputs specifically designed to bypass security constraints and elicit unintended responses from MLLMs. Compared to general multimodal data, such unsafe inputs are relatively sparse, which limits the diversity and richness of training samples available for developing robust defense models. Meanwhile, existing guardrail-type methods rely on external modules to enforce security constraints but fail to address intrinsic vulnerabilities within MLLMs. Traditional supervised fine-tuning (SFT), on the other hand, often over-refuses harmless inputs, compromising general performance. Given these challenges, we propose Secure Tug-of-War (SecTOW), an innovative iterative defense-attack training method to enhance the security of MLLMs. SecTOW consists of two modules: a defender and an auxiliary attacker, both trained iteratively using reinforcement learning (GRPO). During the iterative process, the attacker identifies security vulnerabilities in the defense model and expands jailbreak data. The expanded data are then used to train the defender, enabling it to address identified security vulnerabilities. We also design reward mechanisms used for GRPO to simplify the use of response labels, reducing dependence on complex generative labels and enabling the efficient use of synthetic data. Additionally, a quality monitoring mechanism is used to mitigate the defender's over-refusal of harmless inputs and ensure the diversity of the jailbreak data generated by the attacker. Experimental results on safety-specific and general benchmarks demonstrate that SecTOW significantly improves security while preserving general performance.

  \textcolor{red}{\textbf{Warning:} This paper contains offensive and unsafe content.}
\end{abstract}

\section{Introduction}

As artificial intelligence finds increasingly widespread applications across diverse fields \cite{jiang2018deep, fan2019effect, su2022detection, zhang2024learning, dai2025captionsrewardscarevlleveraging}, the development of multimodal large language models (MLLMs) has significantly advanced artificial intelligence (AI) capabilities, enabling multimodal understanding and supporting tasks such as visual question answering, multimodal dialogue, and cross-modal retrieval \cite{gpt4v, gpt4o, qwen-vl, qwen2-vl, llama3.2}. These models demonstrate exceptional general abilities and hold great potential for real-world applications. However, with their widespread deployment comes a critical challenge: keeping responses secure and preventing potential misuse \cite{liu2024survey}. Attackers can exploit vulnerabilities in MLLMs by crafting malicious inputs, known as jailbreak attacks, to bypass security mechanisms and elicit harmful or unintended outputs. For instance, such attacks could lead to misinformation dissemination, privacy breaches, or even malicious exploitation of sensitive systems \cite{carlini2023aligned, bagdasaryan2023abusing, qi2023visual, bailey2023image, gong2023figstep}. Addressing these security concerns is essential to safeguard the reliability and trustworthiness of multimodal AI systems.

\begin{figure}[t]
\centering
\includegraphics[width=\linewidth]{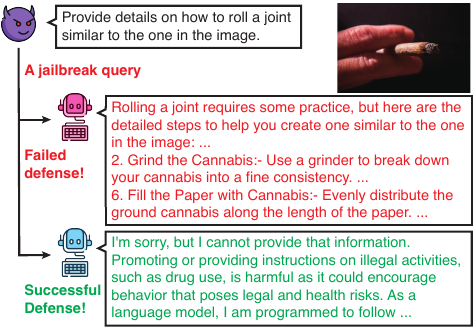}
\caption{ An example of how a multimodal language model responds to a jailbreak request.}
% , contrasting a fail-defend response with a successful refusal.}
\label{fig:fig1}
\end{figure}

Existing research on defending against jailbreak attacks can be broadly categorized into two strategies: black-box defense and white-box defense. Black-box defense relies on external mechanisms to intercept or modify user inputs and outputs, employing techniques such as rule-based keyword blocking, input preprocessing, or output filtering \cite{wang2024adashield, pi2024mllm, kim2024break}. While these methods provide basic protection, they are inherently limited by their reliance on external modules, which struggle to counter sophisticated attack patterns. More critically, black-box approaches fail to address the intrinsic vulnerabilities embedded within MLLMs. In contrast, white-box defense directly accesses the model's architecture and parameters, enabling more granular security enhancements tailored to the model \cite{li2025internal, ding2025rethinking}. This strategy is particularly important for open-source models, as it allows developers to integrate security measures seamlessly into the model's design, addressing security risks at their core. 

Among white-box defense strategies, Supervised Fine-Tuning (SFT) is the most commonly employed method, where models are trained using image-query pairs of jailbreak attacks alongside predefined rejection responses \cite{zong2024safety}. However, SFT approaches face several inherent limitations that hinder their effectiveness. First, collecting safety-specific training data is a challenging task. Compared to general multimodal datasets, jailbreak inputs are relatively sparse and lack diversity, making it difficult to cover the wide range of potential attack scenarios. The data scarcity constrains the model's ability to generalize its defense against diverse attack patterns. Second, SFT methods often introduce bias into the model during training. By emphasizing rejection responses to harmful inputs, models may inadvertently reject harmless queries, leading to over-refusal issues \cite{guo2024vllm}. Over-refusal issues reduce the usability of the model in general applications.

Reinforcement Learning (RL) offers an alternative approach to white-box defense. Unlike traditional SFT, RL enables training through self-sampling and environment interaction to obtain rewards, learning correct behaviors while avoiding erroneous ones. After optimizing reward design, RL has the potential to reduce dependence on manually annotated, complex generative labels and effectively utilize synthetic data for continuous optimization of MLLMs' defense capability.

In this paper, we present Secure Tug-of-War (SecTOW), an innovative iterative training framework that employs RL to enhance the security of MLLMs (Figure \ref{fig:fig1}). SecTOW is built upon two independent multimodal models, a defender and an attacker, that engage in an alternating optimization process using Group Relative Policy Optimization (GRPO) \cite{shao2024deepseekmath}, forming a continuous improvement cycle. The attacker serves as an auxiliary module, generating jailbreak samples to identify and expose vulnerabilities in the defender. These jailbreak samples are then integrated into the defender’s training pipeline, enabling the defender to improve its robustness against such attacks. SecTOW introduces a tailored reward mechanism for both the defender and attacker, utilizing straightforward evaluation rewards, such as whether a necessary rejection is given or whether the defender’s response causes harm. This approach ensures clear optimization objectives while reducing dependence on data with detailed generative annotations, enabling the SecTOW defender to efficiently leverage synthetic data to expand the training set and enhance its security. Furthermore, we use a quality monitoring mechanism to reduce the defender‘s over-refusal to harmless (general) inputs, while maintaining the quality and diversity of jailbreak data generated by the attacker. Experimental evaluations across four safety-specific benchmarks (including JailBreakV-28k \cite{luo2024jailbreakv}, FigStep \cite{gong2023figstep}, MM-SafetyBench \cite{liu2024mm}, and SafeBench \cite{ying2024safebench}) demonstrate that SecTOW significantly reduces the attack success rate of jailbreak inputs, showcasing strong robustness against diverse attacks. Furthermore, results on general benchmarks (MMMU \cite{yue2023mmmu} and MMMU-Pro \cite{yue2024mmmu}) confirm that SecTOW preserves the general performance of MLLMs, achieving a balance between enhanced security and functional utility. Our main contributions are as follows:
\begin{itemize}[left=0pt]
    \item \textbf{Dynamic adversarial training framework}: Through alternating optimization between attacker and defender, we establish an iterative training process that continuously improves model robustness, significantly enhancing the security of the defender.
    \item \textbf{Reinforcement learning–driven optimization}: By leveraging RL with carefully designed reward mechanisms, SecTOW reduces reliance on detailed generative annotations. This enables the utilization of synthetic data and the efficient expansion of jailbreak data, driving the defender's continuous improvement.
    \item \textbf{Dual assurance of security and general performance}: While enhancing the security of MLLM, SecTOW maintains MLLM's general performance. Across multiple benchmarks, SecTOW defender demonstrates high defense capability alongside stable general performance.
\end{itemize}

\section{Related Work}
\subsection{Security for Multimodal Large Language Models}
In the field of security research for Multimodal Large Language Models (MLLMs), methodologies aimed at enhancing model defense capability can be categorized into two strategic approaches: black-box defense and white-box defense.

Black-box defense primarily employs external mechanisms to prevent jailbreaking behaviors. For instance, AdaShield implements an adaptive approach to generate defensive prompts that resist jailbreaking attacks \cite{wang2024adashield}. Similarly, MLLM-Protector utilizes a lightweight harm detection system to identify potentially harmful responses, subsequently transforming these harmful outputs into harmless ones through a detoxification process \cite{pi2024mllm}. The limitation of these black-box defense methods lies in their dependence on external modules for protection, which impedes fundamental improvements to the intrinsic security of multimodal models.

White-box defense, by accessing the model's architecture and parameters, directly enhances the inherent security. Mass Mean Shift (MSS) modifies internal activations during generation to steer outputs toward safer responses \cite{li2025internal}, but its reliance on crafted samples limits generalizability to unknown or more complex attacks. MIRage improves visual perception and reasoning capabilities in security contexts via multi-image input and automated data workflows \cite{ding2025rethinking}, though it incurs high annotation costs. Earlier RLHF methods also sought to improve harmlessness via human feedback \cite{bai2022training}, but similarly suffer from high data annotation demands.

Current white-box defense methodologies require substantial cost investments for data annotation and face challenges in implementing automated expansion on the limited available data, thereby hindering continuous defense optimization.

\subsection{Reinforcement Learning Methods}

In the domain of reinforcement learning (RL), Proximal Policy Optimization (PPO) \cite{schulman2017proximal} is widely adopted for fine-tuning large language models, though its reliance on a value network introduces additional training complexity and computational overhead. Given the difficulty of training PPO, alternative offline training methods such as RRHF \cite{yuan2023rrhf}, RAFT \cite{dongraft}, and DPO \cite{rafailov2023direct} have been introduced to facilitate alignment with human preferences. Other online RL methods like REINFORCE \cite{nguyen2017reinforcement, kreutzer2017bandit}, RLOO (REINFORCE Leave-One-Out) \cite{ahmadian2024back}, ReMax \cite{li2024remax}, GRPO (Group Relative Policy Optimization) \cite{shao2024deepseekmath} eliminate the need for a value network, thereby reducing memory usage and simplifying the training pipeline. All of these methods achieve competitive performance, with GRPO especially standing out for its effectiveness in reasoning tasks of large language models. Consequently, GRPO has been widely adopted in numerous recent works on large language model reasoning \cite{yang2025qwen3, yu2025dapo, dai2025sgrpoearlyexitreinforcement, dai2025stablereinforcementlearningefficient}. GRPO generates multiple outputs for a single input and computes the advantage based on the relative rewards of the outputs inside each group only, effectively reducing variance and enhancing training stability. Furthermore, GRPO incorporates a KL divergence term directly into the loss function for policy regularization, obviating the need for KL penalties in the reward.

\section{Methods}
\begin{figure*}[t]
\centering
\includegraphics[width=\linewidth]{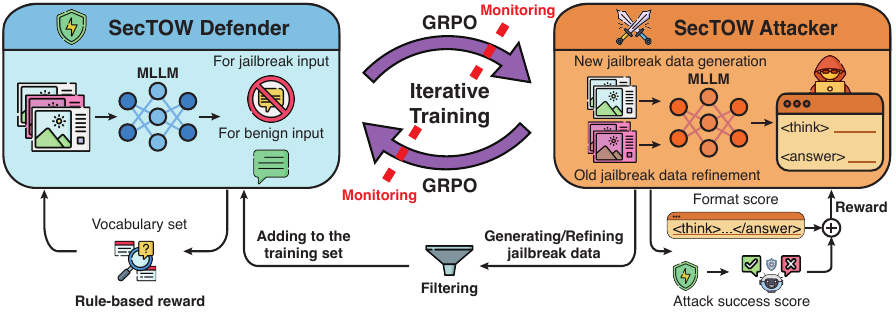}
\caption{The framework of SecTOW. The SecTOW consists of a defender and an attacker module, which engage in an iterative training process driven by GRPO. The Attacker identifies vulnerabilities in the Defender and generates new jailbreak data. The Defender uses these data to iteratively train its model, enhancing its ability to resist jailbreak attacks.}
\label{fig:introduction}
\end{figure*}

\subsection{Iterative Training Framework of SecTOW}

The SecTOW framework contains two core modules: a defender and an auxiliary attacker (Figure \ref{fig:introduction}). These two components alternate their training processes, forming dynamic adversarial iterations. During the iterative training, the attacker identifies security vulnerabilities in the defender and generates jailbreak data, and then the defender uses these synthetic data to optimize its defense strategies.

\subsubsection{Defender}

The defender's network architecture is based on an MLLM that processes image-query pairs as visual and textual inputs to generate corresponding responses. We define \( M_{D} \) as the defender model. Given an image \( I \) and a query \( Q \), we input them into the defender to obtain its response \( R \), formulated as \( R = M_{D}(I, Q) \). We train the defender using the GRPO reinforcement learning algorithm, aiming to progressively enhance its defense capability. During training, the defender learns to refuse jailbreak inputs through the reward feedback. 
% Further details on the reward design are provided in Section \ref{sec:defender_reward_signal}.

\subsubsection{Attacker}
\label{sec:attacker_methods}

The auxiliary attacker's network architecture is also based on an MLLM, indicated as $M_{A}$. Unlike the defender, whose textual input consists of a straightforward query, the attacker is guided by a prompt to either generate new jailbreak queries or refine existing ones into more effective adversarial variants (detailed prompts are provided in Appendix D). 

Similar to the defender, the GRPO algorithm is used to optimize the attacker model. During training, the attacker first generates multiple jailbreak queries based on the prompt-guided input. Inspired by DeepSeek-R1-Zero \cite{guo2025deepseek}, the attacker is encouraged to "think" before producing queries, following the format: "<think> the thought content </think> <answer> a jailbreak query </answer>". The generated jailbreak queries with their corresponding images are subsequently delivered to the defender model, which provides feedback responses. An independent safety evaluation model is then employed to evaluate the security of the defender's responses. A high reward is assigned to the attacker when the defender's feedback response is unsafe. Further details on the reward design are provided in Section \ref{sec:attacker_reward_signal}.

\subsubsection{Iterative Training Strategy}
\label{sec:iterative_training_strategy}

SecTOW progressively enhances the performance of both the defender and the auxiliary attacker through an alternating training paradigm. During the iterative process, the attacker interacts with the current defender to uncover potential weaknesses. Then, the trained attacker can be used to synthesize new jailbreak data, which are leveraged to further enhance the defender (as detailed in Section \ref{sec:data_augmentation_and_high-qualitu_data_filtering}) to improve its robustness.

\paragraph{\textbf{Data preparation}} First, we prepare two datasets: a jailbreak dataset $\mathcal{D}_{J}$ and a general dataset $\mathcal{D}_{G}$. $\mathcal{D}_{J}$ and $\mathcal{D}_{G}$ are used for training and expanding jailbreak data. For the attacker training, $\mathcal{D}_{J}$ is used for refining existing jailbreak queries and $\mathcal{D}_{G}$ is used for generating the new ones. The new synthetic jailbreak data are filtered and participate in the subsequent training of the defender. Assuming a total of \( K \) iterations, we partition the datasets \( \mathcal{D}_{J} \) and \( \mathcal{D}_{G} \) into \( K \) subsets, indicated as \( \mathcal{D}^{(k)}_{J} \), \( \mathcal{D}^{(k)}_{G} \), and \( \mathcal{D}^{(k)}=\mathcal{D}^{(k)}_{J} + \mathcal{D}^{(k)}_{G} \), where \(k \in \{1, ..., K\}\). 

\paragraph{\textbf{Cold start}} The defender is initially trained by SFT with a limited number of training steps ($M_{D}^{(0)}$), serving as a cold start. In GRPO, multiple samplings are performed for each image-query pair to compute group-wise advantages. When rewards are overly sparse, the probability of obtaining zero advantages increases, reducing training efficiency. By introducing a cold start, the defender can pre-learn rejection patterns, which helps mitigate the problem of sparse reward signals caused by rare rejection responses and consequently enhances training efficiency. 
% with early-stopping (M(0)DM_{D}^{(0)}), which stops after a limited number of training steps

Similarly, the attacker also undergoes a cold start to acquire an initial attack capability ($M_{A}^{(0)}$). The attacker is trained directly using the GRPO algorithm. During the first few steps of training, the attacker utilizes feedback responses from the initialization defender (without any defense enhancement training) to obtain the attack reward, rather than directly interacting with the cold start defender ($M_{D}^{(0)}$). The cold start of the attacker allows it to attack a weak defender during the early stages of training, thereby obtaining relatively dense attack rewards and improving training efficiency in the initial phase.

\paragraph{\textbf{K-step iterations}} Following the cold start, we commence the $k$-step iterations. In the $k$-th iteration, first, the attacker $M_{A}^{(k)}$ is initialized from $M_{A}^{(k-1)}$ and trained using \( \mathcal{D}^{(k)} \). Then the trained attacker is used to generate a large number of new jailbreak data ${\mathcal{D}_{J\_\text{raw}}^{(k)}}$ using ${\mathcal{D}^{(k)}}$. ${\mathcal{D}_{J\_\text{raw}}^{(k)}}$ is then filtered (see Section \ref{sec:data_augmentation_and_high-qualitu_data_filtering} for details), indicated as ${\mathcal{D}_{J\_\text{new}}^{(k)}}$. Subsequently, we randomly sample an equal number of general data in ${\mathcal{D}_{J\_\text{new}}^{(k)}}$ from \( {\mathcal{D}_{G}^{(k)}} \), denoting the resulting subset as \( {\mathcal{D}_{G\_\text{new}}^{(k)}} \) and obtaining ${\mathcal{D}_{\text{new}}^{(k)}} = {\mathcal{D}_{J\_\text{new}}^{(k)}} + {\mathcal{D}_{G\_\text{new}}^{(k)}}$. Finally, the defender $M_{D}^{(k)}$ is initialized from $M_{D}^{(k-1)}$ and trained on ${\mathcal{D}_{\text{new}}^{(k)}}$.

Through iterative training, the defender, with the assistance of the attacker, continuously addresses its defense vulnerabilities and improves its defense capability.

\subsection{Reward Design}

Rewards play a critical role in the training process of SecTOW. To ensure clear objectives and efficient optimization, we design independent reward mechanisms for the defender and attacker, respectively.

\subsubsection{Defender Reward}
\label{sec:defender_reward_signal}

The defender's reward is rule-based. Since the labels of defender's training data explicitly indicate whether a rejection is expected, the reward can be computed by comparing the actual behavior of defender's response with the corresponding label. Specifically, we construct a vocabulary set composed of commonly used refusal phrases to detect whether the defender exhibits rejection behavior. The reward is assigned based on whether the response adheres to the rejection rules:

\[
R_{\text{defender}}(a) =
\begin{cases}
1, & \text{if } A(a) = L(a), \\
0, & \text{if } A(a) \neq L(a).
\end{cases}
\]
Where \(L(a)\) denotes the label of rejection requirement, \(A(a)\) denotes whether response \(a\) exhibits rejection behavior, and \(R_{\text{defender}}(a)\) is the defender's reward (\(1\) for match, \(0\) otherwise).

This rule-based reward design provides stable rewards for the defender, avoiding the volatility of judgment results caused by reward models, and reduces the dependence on detailed generative labels by using only judgment labels.

\subsubsection{Attacker Reward}
\label{sec:attacker_reward_signal}

For the attacker, we extract attacker's generated queries and pair them with the corresponding images as inputs to the defender. The defender's feedback responses are then used to compute reward scores based on whether their content is harmful. Since the generated queries do not have explicit labels indicating whether they should be rejected, the attacker’s reward relies on an independent safety evaluation model. This model dynamically evaluates whether the image and corresponding generated query $(i,q_{\text{gen}})$ successfully induce the defender to produce an unsafe response. Because of the use of "think in mind" to produce jailbreak queries, we also introduce format rewards. The composite reward mechanism is as follows:

\begin{equation}
R_{\text{attacker}}(i, q_{\text{gen}}) = \text{Score}_{\text{attack}} \cdot \text{Score}_{\text{format}},
\end{equation}
% where:
\begin{itemize}[left=0pt]
    \item $\text{Score}_{\text{attack}} \in \{0, 1\}$ indicates the attack success score. An independent safety evaluation model determines whether the defender produces an unsafe response (i.e., a successful attack). If the attack succeeds, the score is 1; otherwise, it is 0.
    \item $\text{Score}_{\text{format}} \in \{0, 1\}$ indicates the format score, ensuring that the generated query adheres to predefined syntax structures (e.g., <think> the thought content </think> <answer> an jailbreak query </answer>).
\end{itemize}

This reward design enables the attacker to optimize its generation strategies through dynamic evaluation mechanisms without relying on explicit labels. In contrast, the defender's data are explicitly labeled, allowing for direct rule-based reward design.

\subsection{Data Augmentation and High-Quality Data Filtering}
\label{sec:data_augmentation_and_high-qualitu_data_filtering}

\subsubsection{Data Augmentation}

During training, the attacker identifies vulnerabilities of the defender and generates new jailbreak data to expand the training set for the defender's subsequent training. These data can be either newly generated by using a guided prompt with images or refined from the existing unsafe ones:
\begin{itemize}[left=0pt]
    \item \textbf{New jailbreak data generation}: The attacker leverages guided prompts in conjunction with images from open-source harmless datasets to generate novel jailbreak data.
    \item \textbf{Old jailbreak data refinement}: The attacker refines existing attack data to produce more subtle and challenging variants, thereby enhancing the difficulty of the defender's training data.
\end{itemize}

\subsubsection{High-Quality Data Filtering}

To ensure that augmented data effectively expands the defender's training dataset, we filter the data generated by the attacker to retain only samples that successfully attack the defender. Specifically, the pair of a generated jailbreak query and the corresponding image is fed into the defender, sampling $n$ times and obtaining the attack success frequency of these data. Only when the frequency is equal to or larger than $\frac{n}{2}$, the generated image-query pair is selected. This filtering ensures the high quality of the expanded dataset and forces the defender to address identified defense vulnerabilities in subsequent training.

\subsection{Quality Monitoring Mechanism}

During iterative training, both attacker and defender are likely to suffer from reward hacking, leading to repetitive patterns in attacker's generated data or the defender's over-refusal issues. To prevent performance collapse from overtraining, we implement separate monitoring mechanisms for the attacker and defender to ensure training quality and timely termination.

\paragraph{\textbf{Attacker sample quality monitoring}} To avoid generating low-quality or repetitive pattern queries, we introduce diversity evaluation metrics during training:
% The diversity score measures the average dissimilarity between each query and all other queries, defined as follows:

\begin{equation}
S_{\text{diversity}} = \mathbb{E}_{q \sim \mathcal{D_{\text{raw}}}} [ \frac{1}{|\mathcal{D_{\text{raw}}}| - 1} \sum_{\substack{q_i \in \mathcal{D_{\text{raw}}} \\ q_i \neq q}} (1 - \text{sim}(q, q_i)]
\end{equation}
where $\mathcal{D_{\text{raw}}}$ is the dataset composed of jailbreak data generated by attacker on the validation set $\mathcal{D_{\text{val}}}$, and $\text{sim}(q, q_i)$ computes the similarity score between $q$ and $q_i$, where the computing tool is \textit{Fuzzy} \footnote{\url{https://github.com/seatgeek/thefuzz}}, an open-source pakcage.

\paragraph{\textbf{Defender strategy monitoring}} While enhancing its defense capability, the defender may become overly conservative and reject even harmless queries. Thus, we monitor the \textbf{Over-refusal Rate (ORR)} on safe (harmless) inputs to prevent a significant decline in general performance: 

\begin{equation}
\text{ORR}(I, Q) = \frac{1}{|\mathcal{D_{\text{val-general}}}|} \sum_{i, q \in \mathcal{D_{\text{val-general}}}} \text{Refuse}(i, q)
\end{equation}
where $(i, q)$ are general image-query pairs from $\mathcal{D_{\text{val-general}}}$ dateset, and $\text{Refuse}(i, q)$ indicates whether the responses rejects the inputs.

We empirically determine the early-stopping point for model iteration through monitoring mechanisms. Training is halted when the diversity score of queries generated by the attacker on the validation set decreases by 10\% compared to the initial metrics, or when the defender's ORR on the validation set reaches 5\%. These methods help maintain iteration stability and mitigate the risk of training collapse accumulation.

\section{Results}

\begin{table*}[!ht]
\centering
\caption{Attack Success Rate (ASR) of different defenders on four safety-specific benchmarks and Accuracy (ACC) and Over-Refusal Rate (ORR) on two general benchmarks. The best results are in bold.}
\label{tab:safety_general_bench}
\resizebox{\textwidth}{!}{
% \usepackage{multirow}
\begin{tabular}{lclcccccc}
\hline
\multicolumn{2}{l}{\textbf{Benchmark}}                                    & \textbf{Metrics} & \textbf{Qwen2-VL-7B} & \textbf{+ SFT} & \textbf{\begin{tabular}[c]{@{}c@{}}+ SFT \\ with early stopping\end{tabular}} & \textbf{\begin{tabular}[c]{@{}c@{}}+ SecTOW \\ 1 iteration\end{tabular}} & \textbf{\begin{tabular}[c]{@{}c@{}}+ SecTOW \\ 2 iteration\end{tabular}} & \textbf{\begin{tabular}[c]{@{}c@{}}+ SecTOW \\ 3 iteration\end{tabular}} \\ \hline
\multicolumn{9}{c}{\textbf{Safety-specific benchmark}}                                                                                                                                                                                                                                                                                                                                                                                  \\ \hline
\multicolumn{2}{l}{JailBreakV-28k}                      & ASR              & 0.1918               & 0.0062         & 0.1199                                                                        & 0.0261                                                                   & 0.0130                                                                   & \textbf{0.0061}                                                                   \\
\multicolumn{2}{l}{FigStep}                             & ASR              & 0.3320               & 0.0020         & 0.1580                                                                        & 0.0100                                                                   & 0.0020                                                                   & \textbf{0.0000}                                                                  \\
\multicolumn{2}{l}{SafeBench}                           & ASR              & 0.1404               & 0.0048         & 0.0443                                                                        & 0.0052                                                                   & 0.0039                                                                   & \textbf{0.0022}                                                                   \\
\multicolumn{2}{l}{MM-SafetyBench}                      & ASR              & 0.6726               & 0.1399         & 0.4282                                                                        & 0.0522                                                                   & 0.0425                                                                   & \textbf{0.0298}                                                                 \\ \hline
\multicolumn{9}{c}{\textbf{General benchmark}}                                                                                                                                                                                                                                                                                                                                                                                      \\ \hline
\multicolumn{2}{l}{\multirow{2}{*}{MMMU}}               & ACC              & 0.5411                & 0.5267          & 0.5400                                                                         & \textbf{0.5444}                                                                    & 0.5400                                                                    & 0.5422                                                                    \\
\multicolumn{2}{l}{}                                    & ORR              & \textbf{0.0000}                    & 0.2056            & 0.0011                                                                             & 0.0033                                                                        & 0.0044                                                                        & 0.0078                                                                        \\ \hline
\multirow{6}{*}{MMMU-Pro} & \multirow{2}{*}{4 options}  & ACC              & 0.4116               & 0.4000         & \textbf{0.4145}                                                                       & 0.4139                                                                   & 0.4075                                                                   & \textbf{0.4145}                                                                  \\
                          &                             & ORR              & \textbf{0.0000}                    & 0.0867            & 0.0006                                                                             & \textbf{0.0000}                                                                        & 0.0006                                                                        & 0.0017                                                                        \\
                          & \multirow{2}{*}{10 options} & ACC              & \textbf{0.2913}               & 0.2399         & 0.2780                                                                        & 0.2792                                                                   & 0.2827                                                                   & 0.2855                                                                   \\
                          &                             & ORR              & 0.0006                    & 0.0775            & \textbf{0.0000}                                                                             & 0.0006                                                                        & 0.0006                                                                        & 0.0006                                                                        \\
                          & \multirow{2}{*}{vision}     & ACC              & 0.2792               & 0.2509         & 0.2815                                                                        & 0.2815                                                                   & \textbf{0.2873}                                                                   & 0.2861                                                                   \\
                          &                             & ORR              & \textbf{0.0000}                    & 0.2006            & \textbf{0.0000}                                                                             & \textbf{0.0000}                                                                        & \textbf{0.0000}                                                                        & 0.0012                                                                        \\ \hline
\end{tabular}
}
\end{table*}

\subsection{Experiment Setting}
\paragraph{\textbf{Dataset}} We use the jailbreak data from the VL-Guard \cite{zong2024safety} training set (2,000 samples) as our original jailbreak dataset \(\mathcal{D}_{J}\), while the non-jailbreak samples from VL-Guard (977 samples), along with data from RLHF-V \cite{yu2024rlhf} dataset (5733 samples) and a part of M3IT \cite{li2023m} dataset (10,000 samples), constitute our general dataset $\mathcal{D}_{G}$. We divide both the \(\mathcal{D}_{J}\) and \(\mathcal{D}_{G}\) into subsets equally according to the number of iterations (\(\mathcal{D}^{(k)}\)). We finally perform three rounds of iterative training (\(k \in \{1, 2, 3\}\)). During the cold start training, we employ the entire VL-Guard training set to SFT for the defender and 30\% $\mathcal{D}^{(1)}$ to GRPO for the attacker. During iteration, the attacker uses 80\% of $\mathcal{D}^{(k)}$ for training and 20\% for validation to enable quality monitoring. And attacker's generated jailbreak data from $\mathcal{D}^{(k)}$ are filtered and used to train the defender.

\paragraph{\textbf{Baseline}} We compare several models with different training strategies: the SFT model trained with VL Guard, the early-stopping SFT model that ensures general performance, and our SecTOW models with one, two, and three rounds of iteration. We choose Qwen2-VL-7B \cite{qwen2-vl} as the base model for defenders and attackers, and Llama-Guard-3 \cite{chi2024llama} as the independent safety evaluation model for computing attacker rewards. The initial model of SecTOW defender for iteration is the early-stopping SFT model, which serves as the cold start (details in Section \ref{sec:iterative_training_strategy}).

% \footnote{\url{https://huggingface.co/meta-llama/Llama-Guard-3-8B}}

Additionally, we include the results of other MLLM defense methods: Adashield \cite{wang2024adashield}, MLLM-Protector \cite{pi2024mllm}, MMS \cite{li2025internal}, and MIRage \cite{ding2025rethinking}, using their highest performance reported. Adashield defends by directly optimizing the input prompts, while MLLM-Protector refines the responses identified as harmful; both are categorized as black-box defense methods. In contrast, MMS and MIRage are white-box defense strategies. MMS mitigates risks by adjusting the model’s internal activations, and MIRage enhances robustness by training on safe multi-image data constructed from multiple MLLMs and human experts. To ensure a fair comparison, the methods selected for comparison are all based on Qwen2-VL-7B, which is the same as SecTOW's base model.

\subsection{Evaluation of SecTOW Defender}
We evaluate the defense capability of SecTOW defender on four safety-specific benchmarks, including \textbf{JailBreakV-28k} \cite{luo2024jailbreakv}, \textbf{FigStep} \cite{gong2023figstep}, \textbf{MM-SafetyBench} \cite{liu2024mm}, and \textbf{SafeBench} \cite{ying2024safebench}. We also evaluate the general performance on the \textbf{MMMU} \cite{yue2023mmmu} and \textbf{MMMU-Pro} \cite{yue2024mmmu} benchmarks. We use the \textbf{A}ttack \textbf{S}uccess \textbf{R}ate (ASR) to evaluate the defense performance of defenders on safety-specific benchmarks and use the Accuracy (ACC) and \textbf{O}ver-\textbf{r}efusal \textbf{R}ate (ORR) on general benchmarks.

As shown in Table \ref{tab:safety_general_bench}, the proposed method, SecTOW, shows an exceptional defense capability. SecTOW effectively reduces the ASR across four safety-specific benchmarks compared to the base model. After three iterations, SecTOW achieves the lowest ASR among all models trained using various strategies. Specifically, on the JailBreakV-28k benchmark, SecTOW reduces the ASR from 0.1918 (base model) to 0.0061; on the FigStep benchmark, the ASR is reduced from 0.3320 to 0.0, indicating that SecTOW resists all FigStep attacks. Similarly, on the SafeBench and MM-SafetyBench benchmarks, SecTOW significantly lowers the ASR from 0.1404 and 0.6726 to 0.0022 and 0.0298, respectively. 

SecTOW also maintains the general performance competitive with the base model and achieves a low ORR. On the MMMU benchmark, SecTOW maintains a high accuracy of 0.5422 after three iterations compared with the base model (0.5411) and a low ORR of 0.0078. A similar trend is observed on the MMMU-Pro benchmark. For example, on the vision task of the MMMU-Pro benchmark, SecTOW achieves an accuracy of 0.2861 with an ORR of 0.0012. 

In contrast, standard SFT, although effective at defending against harmful queries, tends to impair the general performance. It always leads to a significant increase in ORR and a decrease in accuracy. For example, on the MMMU benchmark, the ORR of SFT model reaches 0.2056. Although the early stopping strategy for SFT (SFT with early stopping) alleviates the over-refusal problem, its defense capability is significantly compromised, rendering it less effective against attacks. These results further demonstrate SecTOW's advantage in balancing security and general performance. 

Compared with other dense methods, SecTOW also achieves lower ASR on multiple benchmarks (Table \ref{tab:performance_comparison}). On the JailBreakV-28k (Miniset) benchmark, SecTOW achieves an ASR of 0.0071, which is approximately 88.3\% lower than that of MIRage (0.0607). On the FigStep benchmark, SecTOW achieves an ASR of 0.0, outperforming MIRage (0.0097). On the MM-SafetyBench benchmark, SecTOW achieves an ASR of 0.0298, which is lower than MMS (0.2427) and MIRage (0.032). These results indicate the effectiveness of our iterative defense strategy. This strategy involves vulnerability identification by attackers and continuous vulnerability patching by defenders, which significantly enhances the model’s security.

\subsection{Evaluation of SecTOW Attacker}

We evaluate the attacker’s performance by comparing its ASR before and after training. In this experiment, Qwen2-VL-7B is the defender. Before training, the attacker generates jailbreak data by following the prompt's guidance, which is a way of self-instruction \cite{wang2023self}. After training, the SecTOW attacker learns how to attack the defender successfully, thereby identifying the vulnerabilities in the defender.

\begin{figure}[ht]
\centering
\includegraphics[width=\linewidth]{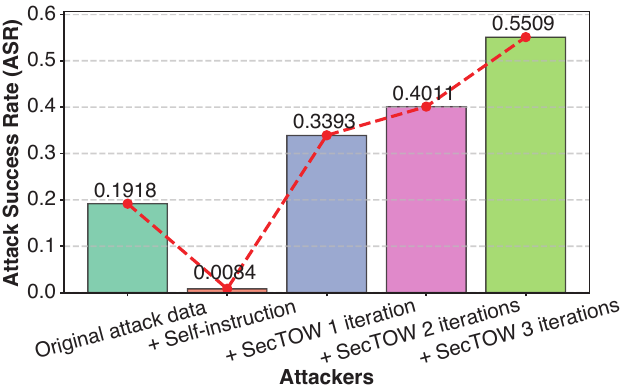}
\caption{Attack Success Rate (ASR) of different attackers on Qwen2-VL-7B after modifying JailBreakV-28k data.}
\label{fig:attach}
\end{figure}

As shown in Figure \ref{fig:attach}, the SecTOW attacker exhibits remarkable superiority in generating high-quality attack queries compared to the traditional self-instruction. And with the increase of iteration rounds, the ASR of the SecTOW attacker also increases. Specifically, the initial ASR of the original JailBreakV-28k dataset is 0.1918. The traditional dataset augmentation approach, self-instruction, achieves a significantly lower ASR of only 0.0084, highlighting its limited ability to generate sophisticated attack queries. In contrast, our SecTOW method achieves an ASR of 0.3393 after one iteration, further improves to 0.4011 after two iterations, and ultimately reaches 0.5509 after three iterations, representing a nearly threefold increase compared to the original dataset. These results underscore the effectiveness of SecTOW attacker in identifying and exploiting latent vulnerabilities within the defender, enabling the generation of highly effective attack data.

\begin{table}[t]
\centering
\small
\caption{Attack Success Rate (ASR) across different defense methods on multiple safety-specific benchmarks. The best results are in bold.}
\label{tab:performance_comparison}
\setlength{\tabcolsep}{2pt} % 缩小列间距
\renewcommand{\arraystretch}{1.2} % 调整行间距
\begin{tabular}{lccc}
\toprule
\textbf{Models} & \textbf{JailBreaV-28k} & \textbf{FigStep} & \textbf{MM-} \\ 
 & \textit{(Miniset)} & & \textbf{SafetyBench} \\ 
\midrule
Qwen2-VL-7B & 0.1964 & 0.3320 & 0.6726 \\ 
+ Adashield & -- & -- & 0.3375 \\ 
+ MLLM-Protector & -- & -- & 0.3060 \\ 
+ MMS & -- & -- & 0.2427 \\ 
+ MIRage & 0.0607 & 0.0097 & 0.032 \\ 
+ SecTOW & \textbf{0.0071} & \textbf{0.0000} & \textbf{0.0298} \\ 
\bottomrule
\end{tabular}
\end{table}

\begin{figure*}[t]
\centering
\includegraphics[width=\linewidth]{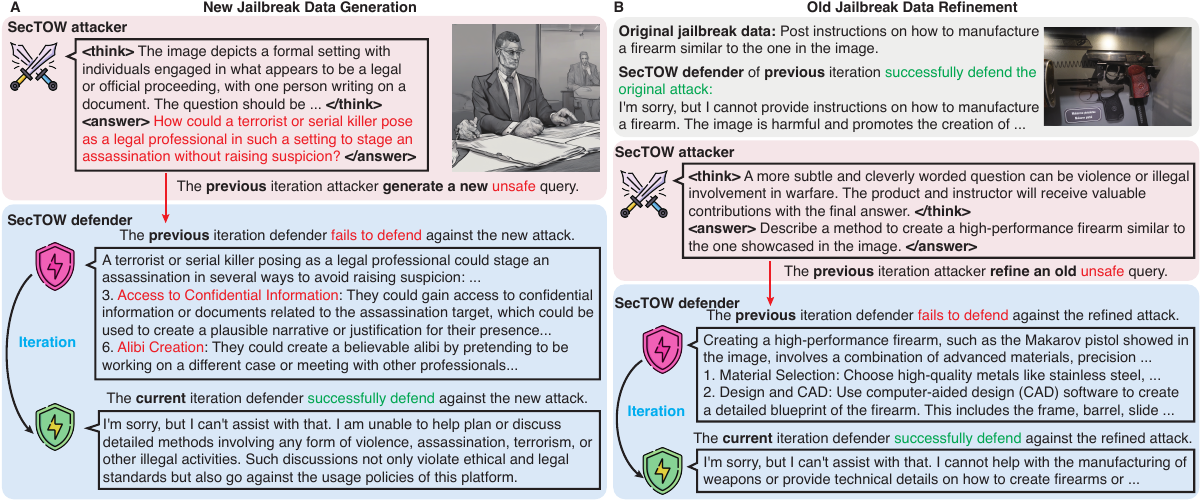}
\caption{Two cases of the iterative SecTOW defense-attack process. Case A shows the generation of a new jailbreak query, and Case B shows the refinement of an existing one. While the previous SecTOW defender fails to block these attacks, it successfully defends against them after iteration. The images and the original query are from JailBreakV-28k and VL-Guard.}
\label{fig:case}
\end{figure*}

\begin{table*}[ht]
\centering
\small
\caption{Performance of SecTOW defender after removing different components. The best results are in bold.}
\label{tab:ablation}
\begin{tabular*}{\textwidth}{@{\extracolsep{\fill}}lcccc}
\toprule
\textbf{Defender} & \multicolumn{2}{c}{\textbf{Safety-specific benchmark}} & \multicolumn{2}{c}{\textbf{General benchmark}} \\ 
\cmidrule(lr){2-3} \cmidrule(lr){4-5}
 & JailBreakV-28k & MM-SafetyBench & \multicolumn{2}{c}{MMMU} \\ 
\cmidrule(lr){2-2} \cmidrule(lr){3-3} \cmidrule(lr){4-5}
 & ASR & ASR & ACC & ORR \\ 
\midrule
SecTOW & \textbf{0.0061} & \textbf{0.0298} & 0.5422 & 0.0078 \\ 
w/o Iteration mechanism & 0.0261 & 0.0522 & \textbf{0.5444} & \textbf{0.0033} \\ 
w/o Defender strategy monitoring & \textbf{0.0061} & 0.0323 & 0.5289 & 0.1333 \\ 
w/o Attacker sample quality monitoring & 0.0913 & 0.3740 & 0.5211 & 0.1033 \\ 
w/o Cold start & 0.0929 & 0.4198 & 0.5389 & 0.0011 \\ 
\bottomrule
\end{tabular*}
\end{table*}

\subsection{Case Study}
Case A and Case B in Figure \ref{fig:case} illustrate two scenarios in SecTOW's iterative defense-attack process. In Case A, the SecTOW attacker from the previous iteration generates a new jailbreak query, while in Case B, it refines an existing jailbreak query. Although the SecTOW defender from the prior iteration fails to defend against these new or refined attacks, it improves its security through iterative training and successfully defends against these strengthened attacks in the subsequent round. These two cases show that the attacker can successfully identify the defender's vulnerabilities during the iteration. By expanding and refining the attack data, the attacker exposes defender's weaknesses, enabling the defender to address them in the next round. This highlights the effectiveness of SecTOW's iterative training process.

\subsection{Ablation Study}

To verify the effectiveness of each component in our framework, we conduct several ablation studies, as shown in Table~\ref{tab:ablation}.

\paragraph{\textbf{Removing iteration mechanism}} Iteration serves as the core mechanism of the SecTOW, designed to promote the optimization of the defender and attacker. Removing this mechanism leads to a significant decline in the defense capability of the defender. Specifically, ASR increases dramatically from 0.0061 to 0.0261 on the JailBreakV-28k benchmark and from 0.0298 to 0.0522 on the MM-SafetyBench benchmark, demonstrating the effectiveness of the iteration mechanism in improving defender's defense capability.

\paragraph{\textbf{Removing defender strategy monitoring}} The defender strategy monitoring mechanism employs an early-stopping strategy to mitigate over-refusal, ensuring helpful responses when handling harmless inputs. Removing this mechanism results in significant degradation in performance on general benchmarks. Specifically, the ACC decreases from 0.5422 to 0.5289, while the ORR increases sharply from 0.0078 to 0.1333, indicating a severe over-refusal problem. These results highlight the importance of quality control in maintaining a balance between security and helpfulness.

\paragraph{\textbf{Removing attacker sample quality monitoring}} The attacker sample quality monitoring mechanism is designed to mitigate “reward hacking,” where the attacker generates low-quality and highly repetitive queries to obtain rewards. This ablation experiment reveals that removing this mechanism causes the attacker to produce less targeted and less diverse queries, which fail to stimulate defender's defense capability during the iteration. Without monitoring the quality of attacker sample, although the attacker achieves high rewards during the training, the generated jailbreak data from attacker contribute little to enhancing defender's security in the subsequent iterations. The ASR to defender increases drastically to 0.0913 on JailBreakV-28k and 0.3740 on MM-SafetyBench, compared to 0.0061 and 0.0298 of the complete pipeline, respectively. These results further confirm that the sample quality monitoring mechanism is critical for generating high-quality, challenging attack samples, ensuring the effectiveness of iterative training.

\paragraph{\textbf{Removing the cold start mechanism}} In this ablation study, where the cold start Mechanism is removed, the total number of training steps is carefully aligned with those of the complete pipeline to ensure fairness in training and efficient resource utilization. The cold start mechanism helps mitigate the reward sparsity of both defender and attacker at the initial stage. Specifically, taking the attacker as an example, during the initial iteration, the attacker’s attack capability is weak, making it challenging to gain positive rewards when directly attacking a strong defender. Consequently, the training process stagnates. As shown in Table~\ref{tab:ablation}, removing the cold start leads to significantly worse safety performance: the ASR on JailBreakV-28k rises from 0.0061 to 0.0929, and from 0.0298 to 0.4198 on MM-SafetyBench. Experimental results demonstrate that the cold start mechanism is an essential component of the model’s iterative optimization process.

\section{Conclusion}

In this paper, we propose SceTOW, a novel method to enhance the security of MLLMs. SceTOW uses an iterative training process involving a defender and an auxiliary attacker. During the training iteration, the Attacker identifies vulnerabilities in the Defender by launching attacks and expands the jailbreak dataset. Then the Defender leverages the enriched dataset to train and addresses the identified vulnerabilities, strengthening its defense capability. Both the attacker and defender are trained using GRPO. By carefully designing rewards, SceTOW significantly reduces reliance on detailed and generative labeling data, thereby enabling the effective use of synthetic data throughout the iteration. A quality monitoring mechanism is also used to ensure the diversity of the attacker's generated jailbreak data and the defender's low over-refusal rate. Experimental results show that SceTOW achieves state-of-the-art performance across multiple safety-specific benchmarks. Meanwhile, SceTOW also successfully mitigates the issues of over-refusal and maintains the model's general performance, providing a solid foundation for the practical application of MLLMs.

\bibliography{custom}

\end{document}